\documentclass[twocolumn, trackchanges]{aastex701}
\usepackage{CJK}
\usepackage{xcolor}

\newcommand{\WL}[1]{\textcolor{black}{#1}}

\begin{document}
\begin{CJK*}{UTF8}{gbsn}

\title{Tidal Synchronization of Binaries in Pleiades}

\correspondingauthor{Li Wang}
\email{wangli79@mail2.sysu.edu.cn}

\author[orcid=0000-0003-3471-9489]{Li Wang (王莉)}
\affiliation{School of Physics and Astronomy, Sun Yat-sen University, Daxue Road, Zhuhai, 519082, China}
\affiliation{CSST Science Center for the Guangdong-Hong Kong-Macau Greater Bay Area, Zhuhai, 519082, China}
\email{wangli79@mail2.sysu.edu.cn}  

\author[0000-0001-9131-6956]{Chenyu He (贺辰昱)}
\affiliation{School of Physics and Astronomy, Sun Yat-sen University, Daxue Road, Zhuhai, 519082, China}
\affiliation{CSST Science Center for the Guangdong-Hong Kong-Macau Greater Bay Area, Zhuhai, 519082, China}
\email{hechy39@mail.sysu.edu.cn}

\author[0000-0002-3084-5157]{Chengyuan Li (李程远)}
\affiliation{School of Physics and Astronomy, Sun Yat-sen University, Daxue Road, Zhuhai, 519082, China}
\affiliation{CSST Science Center for the Guangdong-Hong Kong-Macau Greater Bay Area, Zhuhai, 519082, China}
\email{lichengy5@mail.sysu.edu.cn}

\author[0000-0001-9313-251X]{Gang Li (李刚)}
\affiliation{Centre for Astrophysics, University of Southern Queensland, Toowoomba, QLD 4350, Australia}
\email{Gang.Li@unisq.edu.au}

\begin{abstract}

Tidal interactions in close binaries play a key role in the long-term rotational and orbital evolution. The distributions of circularization across open clusters (OCs) place strong observational constraints on tidal dissipation in binaries. However, direct observational constraints on synchronization among binaries in OCs remain limited. For the 125 Myr OC Pleiades, this work combines cluster membership from {\sl Gaia Data Release 3}, rotation periods from the {\sl K2} mission, and orbital solutions of the binary population from a long-term spectroscopic survey, to investigate the degree of tidal synchronization in each binary by comparing the pseudo-synchronization period to the rotation period of the primary stars. Among \WL{42} binaries with reliable orbital periods $P_{\rm orb}$ and rotation periods, we identify seven tidally synchronized systems with $P_{\rm orb}\lesssim 8.6$ days, including one early-type system and six late-type systems. For binaries with longer $P_{\rm orb}$, primaries generally are super-synchronized, and most systems are eccentric. We find a synchronization transition near $P_{\rm orb}\approx 8.6-14$ days, comparable to the known circularization period ($P_{\rm orb}\approx 7.2$ days) in the Pleiades, which suggests similar critical period scales for synchronization and circularization in this coeval population. Synchronization depends much more strongly on mass ratio than on primary mass. Most synchronized systems in Pleiades have high mass ratios and are likely to evolve into double white dwarf systems. Tides \WL{likely} impose strong rotational braking on close early-type binaries, while their influence on late-type close binaries is weaker, and their spins largely follow the single-star sequence.

\end{abstract}


\section{INTRODUCTION} 

More than half of all stars reside in binary or multiple systems, making binary evolution a central topic in modern astrophysics \citep{Abt1976}. In close binaries, tidal interactions dissipate angular momentum and energy and drive the system toward a minimum-energy, maximum-entropy equilibrium. The expected outcomes include spin-orbit alignment, synchronization of stellar rotation with orbital motion, and orbital circularization \citep{Hut1980}. This tide-driven evolution links binary physics to stellar population statistics and is key to understanding internal stellar structure, angular-momentum loss processes, and the long-term evolution of binary systems. Accurate measurements of tidal efficiencies and timescales are therefore essential for both theory and observation.

The physics of tidal dissipation in binary stars has been debated for several decades, resulting in two representative theoretical frameworks: equilibrium tides and dynamical tides \citep[see review of][]{Ogilvie2014}. \WL{The equilibrium tides correspond to large-scale non-oscillatory tidal flows.} In this picture,  dissipation in late-type stars arises from turbulent friction acting in convective envelopes, with additional radiative damping relevant for early-type stars \citep[e.g.,][]{Zahn1977, Zahn1989}. This mechanism is efficient in stars that possess outer convective envelopes \WL{\citep{Zahn1989, Zahn1989b}}. By contrast, dynamical tides treat the dissipation of \WL{small-scale} tidally-forced oscillations: internal gravity modes \WL{excited in radiative regions and} damped by radiative diffusion \citep[e.g.,][]{Zahn1975, Zahn1977, Goodman1998, Terquem1998}, and rotationally supported inertial waves \WL{excited in convective envelopes and} damped by viscosity \citep[e.g.,][]{Wu2005, Ogilvie2007, Goodman2009}. Both mechanisms remove orbital energy and angular momentum, leading to secular changes in orbital period, eccentricity, and stellar rotation. Tides generally circularize orbits and drive each star toward pseudo-synchronization between spin and orbital motion \citep{Hut1981}. 
\WL{However, dynamical tides associated with self-excited stellar oscillations do not necessarily lead to synchronization; instead, they can drive inverse tides, in which angular momentum is transferred from the star to the orbit, pushing the system away from spin--orbit synchronization \citep{Fuller2021}. 
Using a direct-solution treatment of dynamical tides implemented in GYRE-tides, \cite{Sun2023} and \cite{Townsend2023} showed that tidal torques in radiative stars change sign over a range of stellar rotation rates, thereby blurring the concept of a unique pseudo-synchronized rotation state.
}

Before the era of time-domain surveys for measuring stellar variability, rotational velocities of stars were primarily obtained through spectroscopy, which measures projected rotation velocities ($v$sin$i$) inferred from broadening of spectral lines. $v$sin$i$ is less suitable for constraining models of tidal evolution, because of the ambiguities introduced by the unknown inclination of the rotation axis and non-rotational line broadening from the secondary \citep{Meibom2006}. The advent of high-precision, long-term photometry, especially {\sl Kepler} \citep{Borucki2010}, {\sl K2} \citep{Howell2014}, and {\sl TESS} \citep{Ricker2015} surveys, greatly expanded the number of rotation-period measurements ($P_{\rm rot}$). By tracking starspot modulation, these missions produced unprecedented homogeneous samples of eclipsing binaries (EBs) for testing tidal theory \citep[e.g.][]{Zimmerman2017}. \citet{Lurie2017} showed that most late-type EBs in the {\sl Kepler} field with orbital period $P_{\rm orb}<10$ days are tidally synchronized, with $P_{\rm rot}\approx P_{\rm orb}$. A systematic search of {\sl TESS} data revealed a similar pattern among low-mass FGKM EBs with \cite{Lurie2017}: a synchronous population with $P_{\rm orb}=P_{\rm rot}$ and a subsynchronous population with $8P_{\rm orb}\approx 7P_{\rm rot}$ \citep{HobsonRitz2025}. \WL{Plausible interpretations for such subsynchronous rotation include differential rotation \citep{Lurie2017}, or a competition between tidal dissipation and magnetic braking \citep{Fleming2019}.}
\WL{Meanwhile, pulsations in some EB systems provide a unique probe of stellar rotation and tidal effects. Using gravity modes, \cite{Li2020} measured near-core rotation rates in early-type stars in EBs and found a clear synchronization trend, together with several systems exhibiting extremely slow rotation. The presence of extremely slowly rotating stars in short-period EBs blurs the boundary of pseudo-synchronization.}

Precise ages, together with the evolutionary state and history of binaries, are essential for measuring the pace of tidal synchronization. Coeval populations with well-determined ages are therefore especially valuable, and young open clusters (OCs) provide natural laboratories with a common age and metallicity but diverse binary properties. The observed distributions of circularization across OCs place strong constraints on tidal dissipation \citep[e.g.,][]{Penev2022, Barker2022, Mirouh2023}. Beyond constraining \WL{the efficiency of tidal dissipation}, tides also shape the rotational evolution of cluster members, which in turn changes stellar surface temperatures and evolutionary lifetimes \citep{DAntona2015}. \citet{DAntona2015} proposed that tidally-synchronized binaries may account for the bimodality of stellar rotation distribution of the split main-sequence (MS) and extended MS turnoff stars in young massive clusters. \WL{$N$-body simulations further suggest that, among the population of MS binaries in clusters, tidal synchronization is predominantly found in systems with near-equal masses \citep{Wang2023}.} Early observational work focused mainly on the relation between the characteristic circularization periods $P_{\rm circ}$ and cluster ages \citep{Meibom2005}. In contrast, direct observational constraints on synchronization among MS binaries in OCs remain limited. \citet{Meibom2006} reported preliminary results for 13 solar-type detached binaries in M35 (NGC\,2168; $\sim$150 Myr) and M34 (NGC\,1039; $\sim$250 Myr), finding subsynchronous rotation in close systems. We still lack a complete picture of how synchronization depends on period and mass and of its overall impact in cluster environments.

The Pleiades is an excellent laboratory for assessing tidal synchronization in cluster binaries. It is nearby, rich in members, and has a well-established age of $\sim$125 Myr with robust membership catalogs \citep[e.g.,][]{Long2023}. Rotation periods across a wide range of spectral types are available from {\sl K2} and {\sl TESS} \citep[e.g.,][]{Long2023}, providing a high signal-to-noise snapshot of how tides interact with stellar spin on the MS at the age of Pleiades. More importantly, more than 43 years of systematic spectroscopic monitoring have greatly refined the census of spectroscopic binaries (SBs) with reliable orbital solutions in the Pleiades \citep{Torres2021}. Taken together, these data combine high-quality spectroscopy with long-term photometry, making the Pleiades an ideal testbed for empirical tests of tidal synchronization and circularization in a coeval population. Here we aim to identify and characterize tidally synchronized binaries in the Pleiades and to compare them with results from $N$-body simulations incorporating stellar evolutionary models.

This paper is organized as follows: Section \ref{sec2} describes the observations and the $N$-body simulation. Section \ref{sec3} presents the main results from the observations and the simulations in detail, followed by a discussion in Section \ref{sec4} and a summary in Section \ref{sec5}.

\section{DATA REDUCTION}\label{sec2}

\subsection{Observations}\label{sec2.1}
\begin{figure*}[ht!]
\epsscale{1.1}
\plottwo{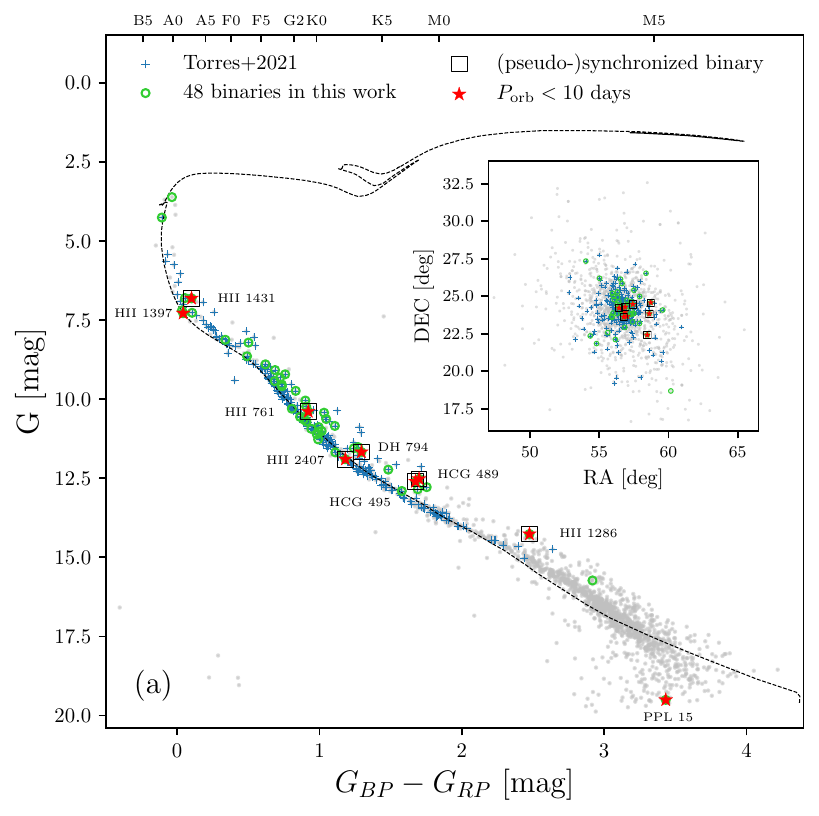}{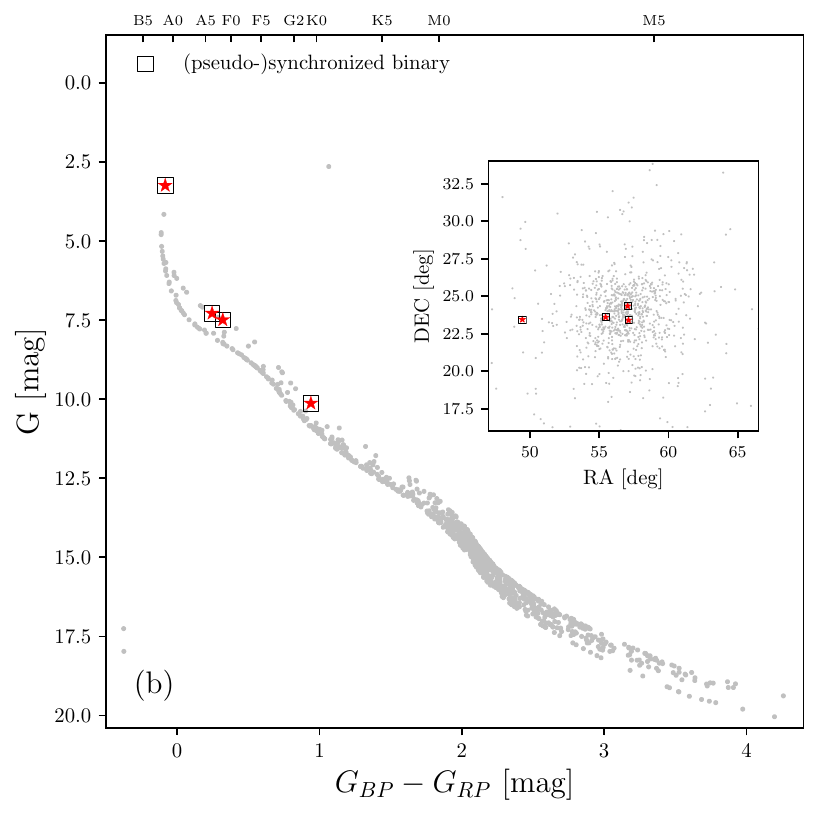}
\caption{(a) CMD and spatial distribution of the Pleiades members. The Black dashed line is a 125 Myr PARSEC isochrone with $Z$ = 0.0152 \citep{Bressan2012}. Interstellar extinction has been applied to the isochrone based on an assumed average reddening of $E(B-V)$ = 0.04. Blue pluses denote Pleiades members from the spectroscopic survey of \cite{Torres2021}. Limegreen open circles represent \WL{48} binaries with reliable orbital solutions from previous literature. Red stars correspond to the close binaries from Table~\ref{t1} with orbital periods $P_{\rm orb}\leq 10$ days, among which the (pseudo-)synchronous systems are highlighted with black open squares. (b) As panel (a), but for the simulated Pleiades-like cluster. We plot only tidally synchronized binaries with $G_{\rm mag}<15$ mag as red star symbols to facilitate comparison with the observations.}
\label{f1}
\end{figure*}

We directly adopt the Pleiades stellar catalog of \cite{Long2023} based on {\sl Gaia Data Release 3} \citep[DR3;][]{Gaia2023}, which provides detailed information on 5D phase-space ($\alpha$, $\delta$, $\mu_{\alpha} \rm{cos}\delta$, $\mu_{\delta}$, parallax) membership probabilities. To ensure a reliable sample, we retain the high-confidence subsample with membership probability $P\geq0.9$ as cluster members (1664 stars). The spatial distribution and the color-magnitude diagram (CMD) of cluster members are shown in Figure~\ref{f1}(a). \citet{Long2023} also measured rotation periods for cool MS stars using {\sl K2} photometry \citep{Howell2014}, spanning an effective-temperature range of 2800$-$6800 K (roughly corresponding to dereddened color $(G_{BP}-G_{RP})_0 \approx 0.5-4$), providing a homogenized set of rotation periods for cluster members. \WL{Additionally, we supplement this dataset with rotation periods derived from {\sl TESS} observations \citep{Gao2025, Fritzewski2025}.}

To investigate tidal evolution of Pleiades binaries, we require precise, homogeneous rotation periods of primaries paired with reliable orbital parameters of binary systems (e.g., orbital period, eccentricity). We therefore rely primarily on the long-baseline spectroscopic survey of \cite{Torres2021}, which combines more than 43 years of spectroscopic observations over an area of $\sim 10^{\circ} \times 10^{\circ}$ across the cluster. This program targets spectral types from mid-B to early-M and reports precise orbital solutions for \WL{38} identified binaries and higher-order multiples, representing the most comprehensive SB dataset for the Pleiades to date. In addition, we also supplement another 9 SBs and 1 EB confirmed in the literature \citep{Basri1999, David2016, Torres2020, Kounkel2021, Gaia2023, Frasca2025, Torres2025}. In total, we collect \WL{48} binaries with secure orbital solutions, of which \WL{42} also have trustworthy $P_{\rm rot}$ measurements. This binary sample is dominated by FGK-type binaries. Focusing on the short-period regime where tides are strongest, we compile systems with $P_{\rm orb} \leq 10$ days in Table~\ref{t1} and highlight them in Figure~\ref{f1}(a). \WL{The remaining systems with $P_{\rm orb} > 10$ days are listed in Table~\ref{t2} in the Appendix.}

\begin{deluxetable*}{lllllllllll}
\tablewidth{0pt}
\tablecaption{List of Nine Detached Binaries with $P_{\rm orb} \leq 10$ days in the Pleiades \label{tab:description}}
\tablehead{
\colhead{N} & \colhead{Name} & \colhead{Gaia DR3 ID} & \colhead{$G_{\rm mag}$} & \colhead{$P_{\rm orb}$} & \colhead{$P_{\rm rot}$} & \colhead{$e$} & \colhead{$q$} & \colhead{$P_{\rm ps}/P_{\rm rot}$} & Label & Ref \\
   & &   & \colhead{(mag)} & \colhead{(day)} & \colhead{(day)} & & 
}
\startdata
1* & HII\,1431 & 66729880383767168 & 6.816  & 2.461127  & 2.90318 & 0      & 0.7018 & 0.84774 & SB2, EB & 1 \\
   &           &                   & 0.003  & 0.000015  & 0.29465 & ...    & 0.0021 & 0.08604 \\
2  & HII\,1397 & 65207709611941376 & 7.287  & 7.345274  & ...     & 0      & ...    & ...     & SB1  & 1 \\
   &           &                   & 0.003  & 0.000018  & ...     & ...    & ...    & ...     \\
3* & HII\,761  & 65276703968959488 & 10.393 & 3.307289  & 3.158   & 0      & 0.698  & 1.04727 & SB2 & 1, 2, 3 \\
   &           &                   & 0.003  & 0.000003  & ...     & ...    & 0.061  & 0.03101 \\
4* & DH\,794   & 63876402894075904 & 11.674 & 5.694369  & 6.338   & 0.0119 & 0.7103 & 0.89921 & SB2 & 1, 3 \\
   &           &                   & 0.003  & 0.000042  & 1.816   & 0.0021 & ...    & 0.25765 \\
5* & HII\,2407 & 66747816167123712 & 11.900 & 7.050477  & 7.247   & 0      & 0.222  & 0.97288 & SB2, EB & 1, 3, 4 \\
   &           &                   & 0.003  & 0.000009  & 1.778   & ...    & 0.033  & 0.23869 \\
6* & HCG\,489  & 65800930496992512 & 12.520 & 3.108737  & 3.054   & 0      & 0.9851 & 1.01792 & SB2 & 1, 3 \\
   &           &                   & 0.003  & 0.000012  & 0.478   & ...    & 0.0034 & 0.15932 \\
7* & HCG\,495  & 66612473157921024 & 12.604 & 8.57662   & 8.881   & 0.1368 & 0.982  & 1.07442 & SB2 & 1, 3 \\
   &           &                   & 0.003  & 0.00053   & 0.532   & 0.0071 & 0.012  & 0.06535 \\
8* & HII\,1286 & 65005262035285888 & 14.265 & 2.46234   & 2.461   & 0      & 0.98   & 1.00054 & SB2 & 2, 3, 5 \\
   &           &                   & 0.003  & 0.00008   & 0.977   & ...    & 0.14   & 0.39721 \\
9  & PPL\,15   & 65000073712701056 & 19.505 & 5.825     & ...     & 0.42   & 0.85   & ...     & SB2 & 6 \\
   &           &                   & 0.005  & 0.3       & ...     & 0.05   & 0.05   & ...     \\
\enddata
\tablecomments{The second line for each system lists the 1$\sigma$ uncertainties. References in the last column are: (1) \cite{Torres2021}; (2) \cite{Kounkel2021}; (3) \cite{Long2023}; (4) \cite{David2015}; (5) \cite{Frasca2025}; (6) \cite{Basri1999}. Objects with asterisks after the numbers are considered as synchronized or pseudo-synchronized binaries.}
\label{t1}
\end{deluxetable*}

Considering only four objects in our \WL{48} binary sample are M-type, the M-type binary subsample is clearly incomplete, so we analyze the completeness level only for binaries brighter than the M-type regime (earlier than M-type). Cross-matching the spectroscopic targets of \citet{Torres2021} with our membership catalog shows that, within $3^{\circ}$ of the cluster center (\WL{46} of our \WL{48} binaries lie in this region) and for $G_{\rm mag}<13$ mag, the spectroscopic monitoring covered about 78\% of members. We adopt a simple rule on the CMD: stars bluer and fainter than \WL{the binary locus with a mass ratio of 0.6, derived from the best-fitting PARSEC isochrone for the Pleiades (125 Myr, $Z$ = 0.0152, $E(B-V)$ = 0.04; \citealt{Bressan2012})}, are treated as singles, while the others are treated as binaries. With this classification, the survey of \citet{Torres2021} shows no strong mass ratio bias, with coverage of singles and binaries of about \WL{72\% and 76\%}, respectively. We take this spectroscopic completeness of members as a proxy for the completeness of our binary sample.

For the brightest close binary listed in Table~\ref{t1}, HII\,1431 (EPIC 211082420), no rotation period has been reported in the literature. HII\,1431 is a double-lined SB (SB2) identified by \cite{Torres2021}. Inspection of its {\sl K2} light curve shows an eclipsing binary with mild ellipsoidal variations but no significant out-of-eclipse starspot modulation, precluding a direct photometric determination of $P_{\rm rot}$. As an alternative, we indirectly estimated its $P_{\rm rot}$ from the projected rotation velocity $v$sin$i$. Using the effective temperatures and projected rotational velocities for the primary and secondary components ($T_{\rm eff,p} = 10450\pm500$K, $T_{\rm eff,s} = 7700\pm600$K; $v{\rm sin}i_{\rm p} = 36\pm3 {\rm km/s}$, $v{\rm sin}i_{\rm s} = 30\pm4 {\rm km/s}$) derived from spectroscopic fitting by \cite{Torres2021}, we inferred the component masses and radii ($M_{\rm p} = 2.36\pm0.18 M_{\odot}, R_{\rm p} = 2.14\pm0.09 R_{\odot}, M_{\rm s} = 1.66\pm0.14 M_{\odot}, R_{\rm s} = 1.71\pm0.15 R_{\odot}$) by interpolating the best-fitting 125 Myr PARSEC isochrone of Pleiades \citep{Bressan2012}. Combining these with the spectroscopic orbital solution for this SB2 from \citet{Torres2021} ($M_{\rm p}{\rm sin}^3 i = 2.14\pm0.01 M_{\rm\odot}$, $M_{\rm s}{\rm sin}^3 i = 1.50\pm0.01 M_{\rm\odot}$), and assuming the system has achieved spin-orbit alignment  ($i_{\rm rot}\simeq i_{\rm orb}$), we derive rotation periods of $2.90\pm0.29$ days and $2.78\pm0.50$ days for the primary and secondary, respectively.

As a diagnostic for the degree of tidal synchronization of binary systems, we use the ratio between the predicted pseudo-synchronization period and the measured stellar rotation period. The pseudo-synchronization period $P_{\rm ps}$ is given by equation (42) of \cite{Hut1981}:
\begin{equation}
P_{\rm ps} = \frac{(1 + 3e^2 + \frac{3}{8}e^4) (1 - e^2)^\frac{3}{2}}{1 + \frac{15}{2}e^2 + \frac{45}{8}e^4 + \frac{5}{16}e^6}P_{\rm orb},  
\label{eq1}
\end{equation}
where $e$ is the eccentricity and $P_{\rm orb}$ is the orbital period. Because the orbital angular velocity varies over an eccentric orbit, exact spin-orbit synchronization is unattainable. Instead, \citet{Hut1981} showed that the system evolves toward a stable equilibrium in which the orbit-averaged tidal torque vanishes. This occurs when the stellar spin angular velocity equals the instantaneous orbital angular velocity at periastron (the pseudo-synchronous state), for which the rotation period equals $P_{\rm ps}$. For a binary with a circular orbit, $P_{\rm ps}=P_{\rm orb}$. $P_{\rm ps}/P_{\rm rot} \approx 1$ thus indicates that the binary approaches synchronization in a circular orbit or pseudo-synchronization in an eccentric orbit.

\subsection{$N$-body Simulation}

To investigate the properties of tidal binary populations within the Pleiades, we run a realistic numerical simulation using the high-performance $N$-body code {\tt\string PETAR}. {\tt\string PETAR} allows us to efficiently mimic the evolution of a massive stellar system containing up to $10^5$ particles with a large fraction of binaries, up to unity \citep{Wang2020, Wang2022}. To accurately follow the dynamical and stellar evolution of both single stars and binary systems, the recently updated single and binary stellar evolution codes, {\tt\string SSE} and {\tt\string BSE} \citep{Tout1997, Hurley2000, Hurley2002, Banerjee2020}, were incorporated in {\tt\string PETAR} to simulate wind mass loss, stellar type changes, mass transfer, and binary mergers. The {\tt\string BSE} tide prescriptions are actually a simplification of the equilibrium tide from \cite{Zahn1970, Zahn1975, Zahn1977, Zahn1989}.

We use the newly updated version of the star cluster initial model generator code {\tt\string MCLUSTER} \citep{Kupper2011, Wang2018} to generate a Pleiades-like cluster. We adopt empirically motivated parameters chosen to reproduce the present-day number-density profile and total mass of the Pleiades. The initial total mass is $1300 M_{\odot}$, and the initial half-mass radius is 3 pc, including all stellar components in 3D space. The cluster metallicity is 0.0152. The initial particle masses were randomly sampled from a Kroupa-like initial mass function \citep[IMF,][]{Kroupa2001} with the mass range of 0.08$-$150 $M_{\odot}$. The 3D positions and velocities of stars were randomly sampled from the Plummer density profile \citep{Aarseth1974}. The adopted simulation model in this work is initialized with a 100\% primordial binary fraction. The initial orbital parameter distributions of binary systems impose essential constraints on forming tidal binaries. Our simulations follow the initial properties of binaries described in \citet{Bellon2017}. Their numerical simulations with these assumptions provide qualitatively similar good descriptions of observations of Galactic-field late-type binaries, globular clusters, and OCs. See more details about initial setups of primordial binary population in our previous work \citep{Wang2023}.

We evolve the synthetic cluster for $\sim$125 Myr to match the current age of the Pleiades and confirm that the final snapshot reproduces the observed 3D number-density profile of the Pleiades. In this Pleiades-like cluster model, the fraction of solar-type stars with $P_{\rm orb}<10^4$ days (the period cutoff used in \citet{Torres2021}) is 28\%, consistent with the $25\pm3$\% measured in the Pleiades by \citet{Torres2021}. For a direct comparison with observations, the theoretical luminosities and temperatures of the simulated stars were then converted into the {\sl Gaia DR3} photometric system using the PARSEC Bolometric Correction ({\tt\string YBC}, \citealt{Chen2019}). Figure~\ref{f1} shows the CMDs of Pleiades and the synthetic Pleiades-like cluster. Because our Pleiades binary sample is severely incomplete for M-type systems, we compare with the real observations only for binaries in the Pleiades-like cluster that have $G_{\rm mag}<13$ mag (earlier than M) and $P_{\rm orb}<10^4$ days. The comparison actually focuses on the statistics of solar-type stars, which dominate the Pleiades binary sample. A detailed comparison between the simulated cluster and the observations appears in Section \ref{sec3}, with further discussion in Section \ref{sec4}.

\section{RESULTS}\label{sec3}

\subsection{The $P_{\rm ps}/P_{\rm rot}-P_{\rm orb}$ Diagram}\label{sec: 3.1}

\begin{figure*}[ht!]
\epsscale{1.1}
\plottwo{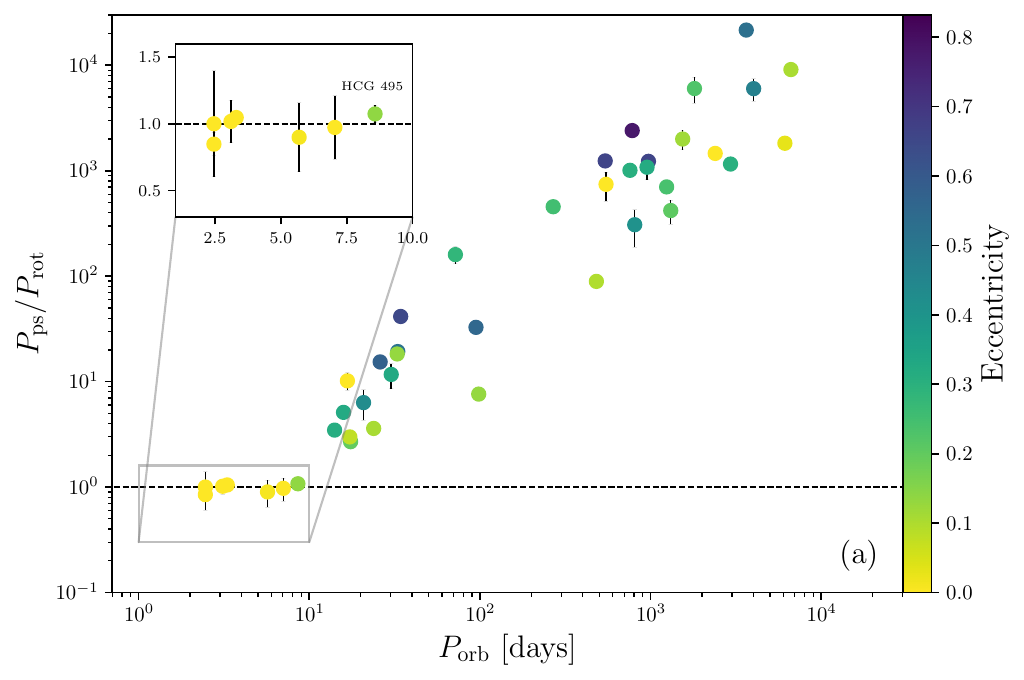}{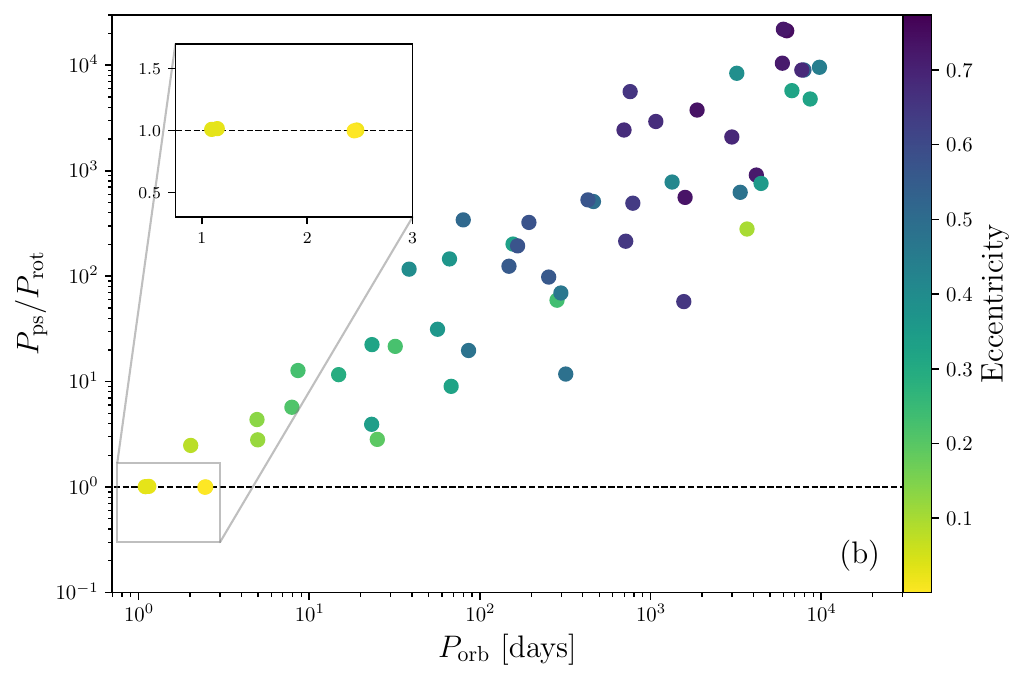}
\caption{(a) Relation between $P_{\rm ps}/P_{\rm rot}$ and $P_{\rm orb}$ for \WL{42} binaries in Pleiades. The colors of the plotting symbols indicate each system's orbital eccentricity. The horizontal black dashed line marks $P_{\rm ps} = P_{\rm rot}$, the synchronization line where systems are synchronized or pseudo-synchronized. The inset highlights the (pseudo-)synchronized binaries. (b) As panel (a), but for the simulated Pleiades-like cluster.}
\label{f2}
\end{figure*}

\begin{figure*}[ht!]
\epsscale{1.1}
\plottwo{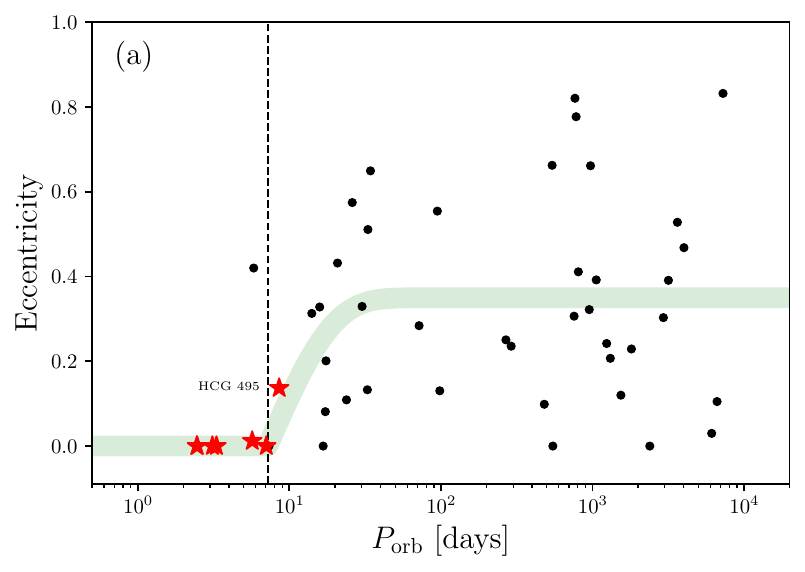}{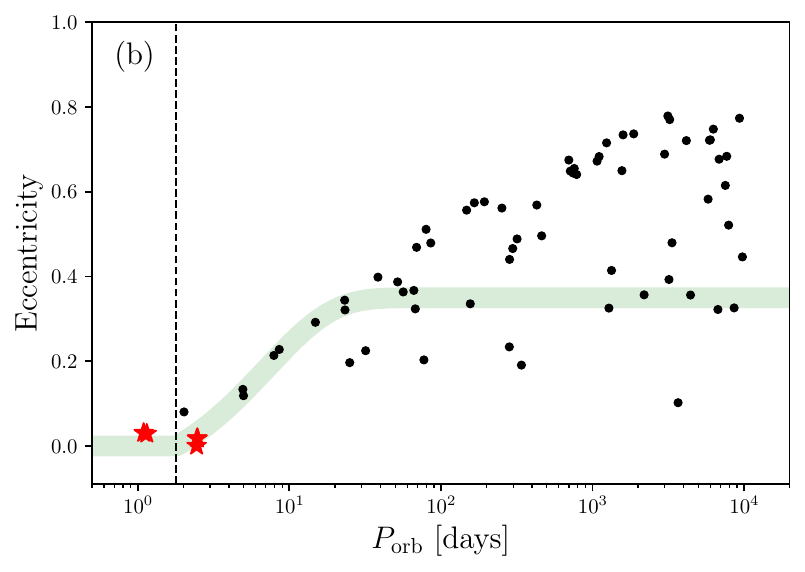}
\caption{(a) Relation between eccentricity and $P_{\rm orb}$ for \WL{42} Pleiades binaries. The green curve represents the circularization function of \cite{Meibom2005} fitted to FGK systems, resulting in a circularization period $P_{\rm circ} = 7.2 \pm 1.0$ days indicated by the black dashed line. Red stars correspond to the (pseudo-)synchronized binaries. (b) As panel (a), but for the simulated Pleiades-like cluster. The resulting circularization period $P_{\rm circ} = 1.8 \pm 0.2$ days.}
\label{f3}
\end{figure*}

The orbital period is the dominant parameter governing tidal synchronization, as the synchronization timescale scales approximately as $P_{\rm orb}^6$ according to \citet{Hut1981}. Figure~\ref{f2}(a) displays the $P_{\rm ps}/P_{\rm rot}-P_{\rm orb}$ diagram for \WL{42} Pleiades binaries. Here, we focus on the degree of tidal synchronization for the following detached binary systems with orbital periods of less than 10 days.

\textit{HII\,1431, HII\,761, DH\,794, HCG\,489, and HII\,1286}: SB2 systems with spectral types A, G, G, M, and M, respectively. The close agreement between each system's orbital period and the primary's rotation period (Table~\ref{t1}) indicates that the primaries are synchronized with their circular orbits, with $P_{\rm ps}/P_{\rm rot} \approx 1$ at the cluster age of 125 Myr.

\textit{HII\,1397}: An A-type single-lined SB (SB1) with a circular orbit of $P_{\rm orb}\sim7.35$ days \citep{Torres2021}. Its companion contributes negligible flux to the system's total light, as indicated by the system's location on the single-star MS in the CMD. \citet{Torres2021} measured projected rotational velocities for objects in the Pleiades field and reported $v\sin i=7\pm3$ km/s for HII\,1397, the slowest for early-type stars with a mean $v\sin i$ near 50 km/s. Even after accounting for inclination, HII\,1397 remains one of the slowest rotators in early-type stars, which likely suggests strong tidal braking in such a close binary. However, no independent rotation period is available, $v\sin i$ alone cannot establish spin-orbit synchronization. We therefore classify HII\,1397 as a candidate near synchronization pending a direct determination of $P_{\rm rot}$ or tighter constraints on the stellar radius and inclination. To further complicate matters, HII\,1397 is part of a complex quadruple system, accompanied by a fainter equal-mass SB2 (HII 1392 \WL{listed in Table~\ref{t2}}) with a 767-day orbital period and a large eccentricity $e$ = 0.82 \citep{Torres2021, Chulkov2025}. Although the {\sl TESS} mission collected high-precision photometry for this source \citep{Ricker2015}, the large pixel size of TESS ($21^{\prime \prime}$) means the photometry is certainly contaminated by flux from nearby sources. The analysis of such a complex, higher-order system is beyond the scope of this paper, which focuses on individual detached binaries.

\textit{HII\,2407}: Previously identified as a G-type SB1 by \citet{Mermilliod1992}, this system was reclassified as an EB system from K2 photometry \citep{Howell2014}. A joint fit of radial velocity and photometric data yielded a full orbital solution and constraints on the fundamental stellar parameters \citep{David2015}. The primary has reached the MS, while the secondary is still on the pre-main sequence (PMS) with a low mass ratio of $q \sim 0.22$, which explains the system's location on the single-star MS in the CMD. The {\sl K2} light curve exhibits starspot modulation, which manifests as roughly sinusoidal variations caused by periodic dips in brightness as spots rotate into and out of view. Based on the phase and amplitude evolution of this modulation, \cite{Long2023} inferred a rotation period of 7.25 days for the primary. Given the circular orbit with $P_{\rm orb}\sim7.05$ days, the primary is synchronized with the orbit, with $P_{\rm ps}/P_{\rm rot}\approx 1$.

\textit{HCG\,495}: An M-type SB2 with a slightly eccentric orbit ($e \approx 0.14$) and $P_{\rm orb} = 8.58$ days, which is the longest period in Table~\ref{t1} \citep{Torres2021}. The primary rotates with $P_{\rm rot}=8.88$ days \citep{Long2023}, close to the pseudo-synchronous period expected for this eccentricity, indicating the pseudo-synchronization state ($P_{\rm ps}/P_{\rm rot} \approx 1$).

\begin{figure*}[ht!]
\epsscale{1.1}
\plottwo{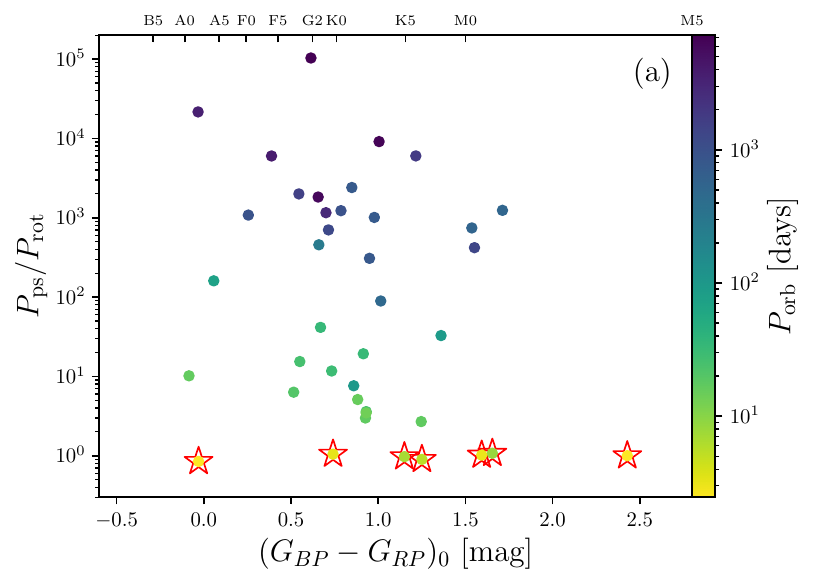}{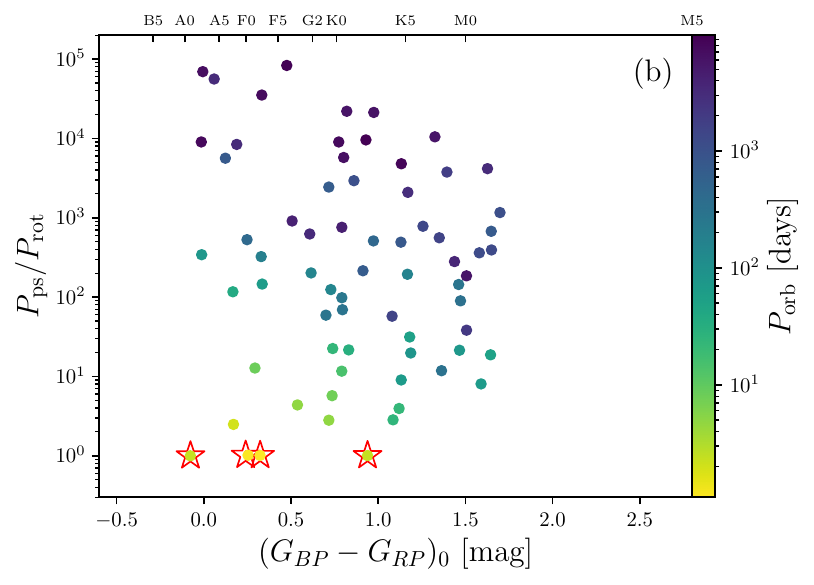}
\caption{(a) Diagram of $P_{\rm ps}/P_{\rm rot}$ versus $(G_{BP}-G_{RP})_0$. Points are color-coded by their $P_{\rm orb}$. Red star symbols mark (pseudo-)synchronized systems. (b) As panel (a), but for the simulated Pleiades-like cluster.
}
\label{f4}
\end{figure*}

\textit{PPL\,15}: This system is the first-known brown dwarf SB reported by \cite{Basri1999}. It has an eccentric orbit ($e \approx 0.4$) with a period of approximately 5.83 days. \cite{Basri1999} found that both components appear to have the same $v$sin$i$ of approximately 10km/s. This is among the lowest rotational velocities observed for very low mass stars in the Pleiades (even after correction for inclination), where rotational velocities of 25-50 km/s are typical. One possibility is that the components have been tidally pseudo-synchronized, meaning their rotational angular velocity matches the orbital angular velocity of the companion at periastron.

Except for HII\,1397 and PPL\,15, for which the primary rotation periods are unknown, the binaries in Table~\ref{t1} with $P_{\rm orb} < 10$ days follow a clear and similar pattern. Primaries in circular orbits are synchronized with the orbital motion ($P_{\rm ps}/P_{\rm rot} \approx 1$). The only exception is HCG\,495, which has the longest orbital period in this subset (8.58 days), a slight eccentricity, and is consistent with pseudo-synchronization. In contrast, systems with $P_{\rm orb} > 10$ days deviate markedly from the synchronization line in Figure~\ref{f2}(a), indicating that their primaries are super-synchronized. Moreover, $P_{\rm ps}/P_{\rm rot}$ increases approximately linearly with $P_{\rm orb}$, which suggests that this ratio is set mainly by the orbital period and tidal torques have a diminishing influence on $P_{\rm rot}$.

In Figure~\ref{f3}(a), the transition period between circular and eccentric orbits for FGK binaries occurs at $P_{\rm orb}\approx 7.2\pm1.0$ days \citep{Torres2021}. Notably, Figures~\ref{f2}(a) and \ref{f3}(a) show that the transition from pseudo-synchronized to non-pseudo-synchronized systems falls at nearly the same orbital period. The existence of HCG\,495 suggests that the pseudo-synchronization boundary ($P_{\rm syn}$) lies at a slightly longer orbital period than the circularization transition ($P_{\rm \mathbf{circ}}$), since it appears pseudo-synchronized while its orbit remains slightly eccentric. It is difficult to confirm precisely due to the lack of observational samples with orbital periods between 8.6 and 14 days. We further find that the $P_{\rm ps}/P_{\rm rot}-P_{\rm orb}$ distribution is tighter and clearer than the $e-P_{\rm orb}$ distribution, and therefore provides a more effective diagnostic of tidal evolution.

As in the observations, four binaries with $0.9 < P_{\rm ps}/P_{\rm rot} < 1.1$ in the simulated Pleiades-like cluster are considered as (pseudo-)synchronous systems and are highlighted with red star symbols shown in Figure~\ref{f1}(b). All these (pseudo-)synchronous binaries are primordial binaries evolving from typical stellar evolutionary processes. These systems are plotted in Figures~\ref{f2}(b) and \ref{f3}(b) for comparison with the Pleiades observations. Overall, the simulated trends in both Figures agree qualitatively with the observation, but the quantitative differences are significant. Tidally synchronized binaries in the simulation concentrate at $P_{\rm orb}\approx 1$ to 3 days. Defining the transition orbital period ($P_{\rm syn}$) between synchronization and pseudo-synchronization as the maximum orbital period among synchronized systems gives $P_{\rm syn}\approx2.5$ days. For the circularization transition, we apply the same procedure as in the observations by fitting the circularization function of \cite{Meibom2005} to the FGK binaries and obtain $P_{\rm circ}=1.8\pm0.2$ days. In this simulation, $P_{\rm circ}$ is close to $P_{\rm syn}$. Both characteristic periods in the Pleiades are clearly larger than these simulated values.

We note that the spectroscopic monitoring completeness of the current binary sample in Pleiades is $\sim$78\%. The impact of undetected systems on the distributions in Figures~\ref{f2}(a) and \ref{f3}(a) remains uncertain. A more complete binary census of the Pleiades is therefore still needed.

\subsection{Dependence on Stellar Mass and Mass Ratio}

We now investigate the dependence of tidal synchronization on stellar mass. For a fixed semi-major axis, the synchronization timescale scales as $R^{-6}$, decreasing with the sixth power of stellar radius \citep{Hut1981}. Larger radii, therefore, lead to faster synchronization and, at a given cluster age, allow synchronization at longer orbital periods. We adopt photometric color as a proxy for binary mass and correct for interstellar reddening using the $E(B-V)$ values for each star from {\sl Gaia DR3} \citep{Gaia2023}. In Figure~\ref{f4}, we compare the distribution of $P_{\rm ps}/P_{\rm rot}$ against dereddened color among the MS binaries. The distributions are similar across these spectral types, and we find no significant dependence of the synchronization state on primary mass within the AFGK regime. The simulated Pleiades-like cluster displays a similar trend.

The synchronization timescale also depends on the mass ratios of binary systems. The tidal synchronization timescale is predicted to decrease with an increasing mass ratio \citep{Hut1981}. Binaries with equal mass ratios ($q = 1$) should have much shorter synchronization timescales than those with low mass ratios, all else being equal. Indeed, we find that synchronization has a stronger dependence on the mass ratio than on the primary mass. In Figure~\ref{f1}(a), of the seven tidally synchronized binaries in the Pleiades, all except HII\,2407 lie noticeably above the observed single-star MS, consistent with their high mass ratios. 86\% (6/7) of these tidally synchronized binaries have $q > 0.6$. We identify high mass-ratio binary candidates directly in the CMD by selecting sources that are brighter and redder than the $q=0.6$ binary locus, and we find that synchronized systems make up about 8\% of this high mass ratio group in Pleiades. \WL{This fraction rises to 17\% when restricted to the spectroscopically observed sample from \citet{Torres2021}, accounting for the incompleteness of spectroscopic coverage.} The FGK SB sample of \citet{Torres2021} is relatively complete and recovers many systems that blend into the single-star MS. As discussed in Section \ref{sec2.1}, their spectroscopic monitoring shows no strong bias with respect to mass ratio. We therefore infer that the higher incidence of synchronized binaries at high $q$ is unlikely to be an artifact of incompleteness. Moreover, in the simulated Pleiades-like cluster, the four tidally synchronized systems we identify are all equal-mass binaries, consistent with the observations as illustrated in Figure~\ref{f1}.

\subsection{Impact of tides on stellar rotation}

\begin{figure}[ht]
\epsscale{1.2}
\plotone{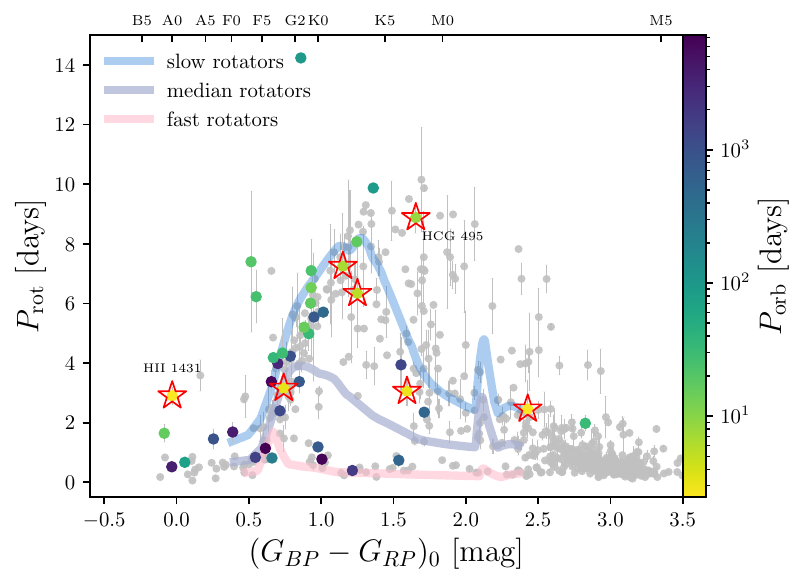}
\caption{Diagram of $P_{\rm rot}$ versus $(G_{BP}-G_{RP})_0$ of Pleiades, highlighting \WL{42} binaries with colors representing their $P_{\rm orb}$. The three curves represent the stellar rotation models of \cite{Amard2019} for low-mass single stars at an age of 125 Myr, with three different initial rotation rates (slow, median, and fast) labeled. Red star symbols and background gray points mark (pseudo-)synchronized binaries and single-star candidates of the Pleiades, respectively.
\label{f5}}
\end{figure}

\WL{To explore how tides shape the rotation of close binaries, we examine their distribution in the $P_{\rm rot}-(G_{BP}-G_{RP})_0$ diagram of the Pleiades (Figure~\ref{f5}).} We compare $P_{\rm rot}$ for \WL{42} binaries with that of single stars, where singles are defined by the renormalized unit weight error (RUWE) parameter in {\sl Gaia} $<$ 1.4 \citep{Lindegren2018}, being photometrically singles on the CMD \WL{as described in Section \ref{sec2.1}}, and not being detected as SB/EBs\WL{, or higher-order systems. We note that this sample of single candidates inevitably contains some unresolved binaries}. Figure~\ref{f5} also overlays the low-mass FGK-type single-star gyrochronology tracks of \citet{Amard2019} for slow, median, and fast initial rotation rates\footnote{The initial rotation rates of slow, median, and fast rotators are 9.0, 4.5, and 1.6(2.3 for the $1.2-1.5 M_{\odot}$ models) days \citep{Amard2019}.} at the cluster age and metallicity. These models implement the wind-braking law proposed by \cite{Matt2015}. 

\WL{For FGK-type single stars, rotation periods primarily fall onto two distinct sequences: a slow-rotator sequence (where Prot increases monotonically with color) and a fast-rotator sequence, with some stars residing in the intermediate gap, as originally demonstrated by \citet{Barnes2003}. This phenomenon has motivated extensive research into the underlying physics \citep[e.g.,][]{Barnes2010, Matt2012, Matt2015, Gallet2015, Gallet2018, Amard2019, Spada2020}}. Late-type tidally synchronized binaries, together with most late-type binaries, fall on the predicted slow-rotator sequence of single stars. We also identify four \WL{long-period binaries ($P_{\rm orb} > 500$ days) that reside on the fast-rotator sequence ($P_{\rm rot} < 1.2$ days). They comprise one young stellar object, two RS Canum Venaticorum variables, and one BY Draconis variable. Early-type singles with $(G_{BP}-G_{RP})_0$ $<$ 0.4 generally rotate rapidly, barring a single exception. Notably, }four of six early-type binaries rotate more slowly than \WL{the bulk of single stars at similar colors.} In this early-type subset, their orbital periods range from 2.46 days (HII\,1431, the synchronized and slowest rotator in this subset) to 4017 days. The primaries appear to rotate more slowly in shorter-period systems, but the small sample does not allow a firm conclusion. 

Above the slow-rotation curve in the \WL{FG-type} regime and the bulk single-star locus in the early-type regime, \WL{five} binaries with relatively short orbital periods, including the synchronized system, HII\,1431, have primaries that rotate anomalously slowly, with rotation periods about 1.3 to 2.5 times longer than those of single stars of similar color. For HII\,1431, this slow rotation further supports a tidal origin for its rotational evolution.

\section{Discussion}\label{sec4}

By directly comparing the pseudo-synchronization period $P_{\rm ps}$ with the primary's rotation period $P_{\rm rot}$, we identify seven tidally synchronized binaries in the Pleiades, comprising one early-type system and six late-type systems. For $P_{\rm orb}<8.6$ days, primaries in detached binaries either reach synchronization on circular orbits or achieve pseudo-synchronization on mildly eccentric orbits ($P_{\rm ps}/P_{\rm rot} \approx 1$). At longer orbital periods, systems are generally super-synchronized and mostly eccentric. Using a coeval, single-metallicity Pleiades sample, we place the synchronization boundary and the circularization boundary in a unified empirical framework and find a key result: the pseudo-synchronization transition ($P_{\rm syn}\approx 8.6$ days) is of the same order as the FGK circularization period ($P_{\rm circ} \approx 7.2\pm1.0$ days), similar to observations in M35 ($\sim$150 Myr) reported by \citet{Meibom2006} and our synthetic Pleiades-like cluster under the classical picture of equilibrium tides \citep{Zahn1989}. 
\WL{Observations of non-cluster binaries from Kepler and TESS data similarly yield synchronization and circularization periods of approximately ten days \citep{Li2020, IJspeert2024}.}
Although the binary sample is sparse for $P_{\rm orb}$ between 8.6 and 14 days, and we do not claim exact equality of the two boundaries, their proximity is unlikely to be a sampling artifact and more plausibly reflects the characteristic scale of tidal dissipation in young coeval populations (125-150 Myr).

In simulation, the Pleiades-like cluster model yields a circularization period for solar-mass binaries of only $P_{\rm circ}=1.8\pm0.2$ days, far shorter than the observed $7.2\pm1.0$ days in cluster Pleiades (Figure~\ref{f3}). The simulated transition from synchronization to pseudo-synchronization also occurs at much shorter orbital periods than observed as shown in Figure~\ref{f2}, which shows that the standard {\tt BSE} tidal prescription (a simplified form of the equilibrium-tide theory \citep{Zahn1970, Zahn1975, Zahn1977, Zahn1989}) is evidently too inefficient to circularize and synchronize binaries with $P_{\rm orb}>2$ days. \citet{Barker2020} argued that classical equilibrium-tide models in MS convective envelopes typically overestimate tidal dissipation by a factor of 2-3, implying that relying on equilibrium tides alone can mislead the interpretation of observations. This motivates revised tidal dissipation physics in which dynamical tides, driven by inertial waves in convective zones and internal gravity waves in radiative zones, should play a key role. \citet{Zanazzi2021} demonstrated that resonant locking of gravity modes during the PMS can drive substantial circularization, with relatively little additional change on the MS. \citet{Barker2022} showed that the observed circularization periods can be explained by dissipation of inertial waves in convective envelopes, which is especially effective during the PMS and can also operate on the MS when the stellar spin approaches synchronization. In this framework, the predicted $P_{\rm circ}$ increases with MS age, consistent with observations. \citet{Barker2022} further demonstrated that neither equilibrium-tide dissipation nor gravity-wave dissipation alone is likely to match the observed $P_{\rm circ}$. Using the large \textit{Gaia} DR3 sample of MS SBs, \citet{Bashi2023} found that circularization is unlikely to occur primarily on the MS and is more consistent with occurring during the PMS, as originally proposed by \citet{Zahn1989}.

HII\,2407 is particularly noteworthy. It is an MS+PMS tidally circularized and synchronized system with a low mass ratio ($q\approx0.22$). Beyond any possible perturbations from an unseen distant tertiary, the primordial eccentricity can modify the time required to reach circularization and synchronization. The circular, synchronized state of this low mass ratio system may therefore reflect an intrinsically small initial eccentricity rather than unusually efficient dissipation. In addition, the secondary remains on the PMS, with a relatively large radius and a deep convective envelope, which would significantly enhance tidal dissipation. HII\,2407 thus potentially supports the view that tidal evolution acts primarily during the PMS and provides useful constraints on the relevant timescales. To refine our understanding of tidal interactions and probe the full potential of tides, we will extend the analysis to clusters spanning a wider range of ages in future work.

The dependence of tidal synchronization on mass ratio is markedly stronger than its dependence on primary mass. Among short-period systems that are (pseudo-)synchronized, high mass-ratio binaries ($q > 0.6$) make up about 86\%, consistent with predictions from our Pleiades-like cluster simulation. The simulation suggests that tidally synchronized binaries in clusters are predominantly ``twin'' binaries, whose fundamental parameters are essentially identical (mass, radius, metallicity, etc.). In the Pleiades, we indeed identify HCG\,489, HCG\,495, and HII\,1286 as equal-mass, tidally synchronized twins. Taken together, the observations and simulations point to a very intriguing conclusion that the tidal-synchronized detached MS binaries in such young clusters ($\sim$125 Myr) are expected to be mostly twin binaries. For the seven synchronized systems in the Pleiades, preliminary {\tt BSE} experiments based on the measured parameters listed in Table~\ref{t1} indicate that HII\,1431 and HCG\,495 are likely to undergo common-envelope evolution and become a double carbon-oxygen white dwarf (WD) system and a double helium WD system, respectively, which are prime targets for next-generation space-based gravitational-wave observatories in the millihertz band \citep{Ren2023}. These results point to star clusters as potential birthplaces of this important class of gravitational-wave sources.

\WL{Evidence that tides shape the rotational evolution of early-type binaries is clear in the Pleiades color-period diagram (Figure~\ref{f5}), albeit limited by the small sample size. Most early-type binaries with $(G_{BP}-G_{RP})_0<0.4$ have primaries that rotate anomalously slowly, consistent with extra tidal braking. The effect strengthens toward shorter orbital periods. The tidally synchronized system HII\,1431 with $P{\rm orb}\approx 2.46$ days hosts the slowest-rotating primary in this subset, whereas systems with $P_{\rm orb} \approx 4017$ days rotate relatively faster. This trend, in which shorter $P_{\rm orb}$ corresponds to slower primary rotation, indicates efficient extraction of spin angular momentum by tides in close binaries. Regarding the sole early-type single candidate with anomalously slow rotation ($P_{\rm rot} \approx 3.6$ days), further investigation is required to determine whether it is an unresolved binary subject to similar tidal interactions. 
In contrast, the rotational distribution of cool single stars in the Pleiades intrinsically exhibits significant scatter. For low-mass binaries, the interplay between magnetic braking and tidal torques makes it even more difficult to isolate the impact of tides on binary rotation. Most late-type binaries lie on the single-star slow-rotator sequence, and their spins do not differ markedly from those of single stars at the same color. This implies that magnetic braking governs the rotational evolution at the present cluster age, with tides contributing negligible additional spin-down or spin-up effects.} 

\section{Summary} \label{sec5}

This work aims to examine the detached tidally synchronized binaries in the cluster Pleiades. Our main results and conclusions are summarized below:
\begin{enumerate}
\item By directly comparing $P_{\rm ps}$ with the primary's rotation period $P_{\rm rot}$, we identify seven tidally synchronized binaries ($P_{\rm orb}\lesssim 8.6$ days) within \WL{42} binaries in the Pleiades, comprising one early-type system and six late-type systems. At longer orbital periods, primaries generally rotate faster than the pseudo-synchronous rate, and most systems are eccentric. 
\item The synchronized system HCG\,495 with the longest orbital period shows a mild eccentric orbit, while the others are circular. We find a synchronization transition near $P_{\rm orb}\approx 8.6-14$ days, comparable to the known circularization period ($P_{\rm orb}\approx 7.2$ days) in the Pleiades, which suggests similar critical period scales for synchronization and circularization in this coeval population. 
\item Synchronization depends much more strongly on mass ratio than on primary mass. Most synchronized systems in Pleiades have high mass ratios. Tidally synchronized detached MS binaries in young clusters ($\sim$125 Myr) are expected to be dominated by twin systems. These high-mass-ratio synchronized binaries are likely to evolve into double WD systems, which are prime targets for next-generation space-based gravitational-wave observatories in the millihertz band \citep{Ren2023}. 
\item \WL{Subject to small-number statistics, our results suggest that} tides \WL{likely} impose strong rotational braking on close early-type binaries, moving them away from the single-star rotation track. In contrast, the tidal influence on late-type close binaries is weaker, and their spins largely follow the single-star slow-rotator sequence.
\end{enumerate}

\begin{acknowledgments}
\WL{We thank the anonymous referee for the valuable comments and suggestions for improving our manuscript.} L.W. and C.L. are supported by the National Natural Science Foundation of China (NSFC) through grant 12233013. 
\WL{C.H. is supported by the NSFC through grant 12503045. G.L. acknowledges the support of the Australian Research Council through the DECRA project DE250100773.} 
This work has made use of data from the European Space Agency (ESA) mission Gaia (https://www.cosmos.esa.int/gaia), processed by the Gaia Data Processing and Analysis Consortium (DPAC, https://www.cosmos.esa.int/web/gaia/dpac/consortium). Funding for the DPAC has been provided by national institutions, in particular the institutions participating in the Gaia Multilateral Agreement.
\end{acknowledgments}

\appendix
\WL{
\startlongtable
\begin{deluxetable*}{lllllllllll}
\tablewidth{0pt}
\tablecaption{List of Detached Binaries with $P_{\rm orb} > 10$ days in the Pleiades \label{tab:description}}
\tablehead{
\colhead{N} & \colhead{Name} & \colhead{Gaia DR3 ID} & \colhead{$G_{\rm mag}$} & \colhead{$P_{\rm orb}$} & \colhead{$P_{\rm rot}$} & \colhead{$e$} & \colhead{$q$} & \colhead{$P_{\rm ps}/P_{\rm rot}$} & Label & Ref \\
   & &   & \colhead{(mag)} & \colhead{(day)} & \colhead{(day)} & & 
}
\startdata
10  & HII\,2168 & 66526127137440128 & 3.616  & 290.992  & ...     & 0.2357 & 0.721  & ...     & SB2  & 7    \\
    &           &                   & 0.004  & 0.003    & ...     & 0.0001 & 0.014  & ...     \\
11  & HII\,563  & 65296907494549120 & 4.261  & 3172.4   & ...     & 0.391  & ...    & ...     & SB1  & 8    \\
    &           &                   & 0.003  & 9.9      & ...     & 0.079  & ...    & ...     \\
12  & HII\,2507 & 66507469798631808 & 6.806  & 16.72623 & 1.645   & 0      & 0.5493 & 10.1679 & SB2 & 1, 3  \\
    &           &                   & 0.003  & 0.00004  & 0.309   & ...    & 0.0007 & 1.90996  \\
13  & TRU\,S194 & 65776736943479808 & 7.197  & 3635     & 0.521   & 0.528  & ...    & 21592.8 & SB1 & 8, 9 \\
    &           &                   & 0.003  & 19       & 0.008   & 0.035  & ...    & 2727.4   \\
14  & TRU\,S26  & 64313222543810560 & 7.271  & 71.820   & 0.671   & 0.284  & ...    & 160.188 & SB1 & 8, 9 \\
    &           &                   & 0.003  & 0.008    & ...     & 0.048  & ...    & 29.1624 \\
15  & HII\,1407 & 64898368889386624 & 8.124  & 953.08   & 1.454   & 0.322  & ...    & 1079.87 & SB1 & 1, 3 \\
    &           &                   & 0.003  & 0.90     & 0.351   & 0.011  & ...    & 262.52   \\
16  & HII\,1762 & 66724451545088128 & 8.214  & 4017     & 1.687   & 0.468  & ...    & 6005.62 & SB1 & 1, 3 \\
    &           &                   & 0.003  & 155      & 0.335   & 0.039  & ...    & 1432.74  \\
17  & TRU\,S93  & 70024154658949120 & 8.646  & 1059.7   & ...     & 0.392  & ...    & ...      & SB1 & 1   \\
    &           &                   & 0.003  & 3.8      & ...     & 0.037  & ...    & ...      \\
18  & HII\,605  & 69819404977607168 & 8.901  & 20.7976  & 7.399	  & 0.4318 & 0.6207 & 6.32416 & SB2 & 1, 3 \\
    &           &                   & 0.003  & 0.0001   & 2.373   & 0.0026 & 0.0068 & 2.02891  \\
19  & AK\,II-346& 64575730944903552 & 9.081  & 6123     & 3.376   & 0.030  & ...    & 1823.48 & SB1 & 1, 3 \\
    &           &                   & 0.003  & 123      & 0.379   & 0.083  & ...    & 214.91  \\
20  &AK\,III-419& 71296667568555904 & 9.215  & 34.32148 & 4.182   & 0.6493 & 0.9003 & 41.4922 & SB2 & 1, 9 \\
    &           &                   & 0.003  & 0.00007  & ...     & 0.0022 & 0.0031 & 1.0194  \\
21  & HII\,745  & 65277975278721152 & 9.320  & 1541.4   & 0.838   & 0.120  & ...    & 1998.56 & SB1 & 1, 3 \\
    &           &                   & 0.003  & 9.8      & 0.171   & 0.025  & ...    & 413.40  \\
22  & HII\,164  & 65090680344356992 & 9.430  & 268.704  & 0.815   & 0.2505 & 0.318  & 455.893 & SB2 & 1, 10 \\
    &           &                   & 0.003  & 0.056    & 0.001   & 0.0072 & 0.017  & 7.485  \\
23  & HII\,233  & 65242069352190976 & 9.540  & 1241.5   & 2.397   & 0.242  & ...    & 702.677 & SB1 & 1, 3 \\
    &           &                   & 0.003  & 1.4      & ...     & 0.025  & ...    & 47.787 \\
24  & HII\,727  & 66802654309459712 & 9.593  & 7271     & 1.139   & 0.832  & ...    & 103223.5 & SB1 & 1, 3 \\
    &           &                   & 0.003  & 153      & 0.314   & 0.039  & ...    & 46833.7   \\
25  & HII\,1392 & 65207709613871744 & 9.740  & 767.04   & ...     & 0.8206 & 0.930  & ...      & SB2 & 1   \\
    &           &                   & 0.003  & 0.25     & ...     & 0.0075 & 0.015  & ...      \\
26  & HII\,1117 & 65199978672758272 & 10.045 & 26.0271  & 6.228   & 0.5745 & 0.980  & 15.3823 & SB2 & 1, 3 \\
    &           &                   & 0.003  & 0.0001   & 0.897   & 0.0062 & 0.014  & 2.2466  \\
27  & HII\,2500 & 66507469798632320 & 10.235 & 2391     & 1.638   & 0      & ...    & 1459.71 & SB1 & 1, 3 \\
    &           &                   & 0.003  & 17       & 0.12    & ...    & ...    & 107.44  \\
28  & HII\,2172 & 66771146429454592 & 10.305 & 30.2130  & 4.337   & 0.3294 & ...    & 11.7012 & SB1 & 1, 3 \\
    &           &                   & 0.003  & 0.0001   & 1.166   & 0.0037 & ...    & 3.1479  \\
29  & HII\,2147 & 66503449709270400 & 10.431 & 6641     & 0.777   & 0.105  & 0.9168 & 9112.99 & SB2 & 3, 11 \\
    &           &                   & 0.003  & 42       & 0.046   & 0.011  & 0.0039 & 555.43  \\
30  & HII\,250  & 69829609819915648 & 10.561 & 971.59   & 4.232   & 0.6613 & ...    & 1228.07 & SB1 & 1, 3  \\
    &           &                   & 0.003  & 0.67     & ...     & 0.0099 & ...    & 64.09   \\
31  & HII\,173  & 69883417170175488 & 10.629 & 481.25   & 5.709   & 0.0986 & 0.953  & 89.2185 & SB2 & 1, 3  \\
    &           &                   & 0.003  & 0.10     & ...     & 0.0067 & 0.011  & 1.6075  \\
32  & HII\,120  & 65232105028172160 & 10.641 & 2940     & 3.985   & 0.303  & ...    & 1157.52 & SB1 & 1, 3  \\
    &           &                   & 0.003  & 12       & 0.385   & 0.026  & ...    & 135.33  \\
33  & HII\,3097 & 66863058730966528 & 10.706 & 780.38   & 3.376   & 0.777  & ...    & 2403.79 & SB1 & 1, 3  \\
    &           &                   & 0.003  & 0.14     & 0.393   & 0.010  & ...    & 326.97  \\
34  & HII\,320  & 68310359628169088 & 10.848 & 757.01   & 1.187   & 0.3064 & 0.857  & 1009.26 & SB2 & 1, 3  \\
    &           &                   & 0.003  & 0.22     & ...     & 0.0049 & 0.023  & 80.49   \\
35  & PELS\,38  & 64743646986745600 & 10.885 & 17.28588 & 6.011   & 0.0811 & 0.573  & 2.98904 & SB2 & 1, 10 \\
    &           &                   & 0.003  & 0.00007  & 0.002   & 0.0018 & 0.012  & 0.00514  \\
36  & HII\,2406 & 64928605459180416 & 10.938 & 33.00630 & 4.986   & 0.5109 & 0.544  & 19.2812 & SB2 & 1, 3  \\
    &           &                   & 0.003  & 0.00005  & 0.572   & 0.0011 & 0.017  & 2.2132  \\
37  & HII\,571  & 69864313155605120 & 11.010 & 15.87225 & 5.202   & 0.3281 & ...    & 5.10767 & SB1 & 1, 3  \\
    &           &                   & 0.003  & 0.00003  & ...     & 0.0031 & ...    & 0.10085 \\
38  &AK\,III-664& 70941383577307392 & 11.023 & 14.073   & 6.531   & 0.313  & ...    & 3.46804 & SB1 & 1, 3  \\
    &           &                   & 0.003  & 0.001    & ...     & 0.011  & ...    & 0.10994 \\
39  & HII\,2284 & 66502281478384000 & 11.115 & 807.39   & 5.545   & 0.4111 & ...    & 307.711 & SB1 & 1, 3 \\
    &           &                   & 0.003  & 0.45     & 2.152   & 0.0061 & ...    & 119.553 \\
40&TYC\,1254-41-1&50176423589328512 & 11.274 & 98.356   & 14.238  & 0.1304 & ...    & 7.61414 & SB1 & 3, 12  \\ 
    &           &                   & 0.003  & 0.427    & 0.118   & 0.0478 & ...    & 0.52380 \\
41  &AK\,I-2-288& 70049477786093056 & 11.516 & 17.4667  & 8.069   & 0.2010 & 0.8700 & 2.69363 & SB2 & 1, 3  \\
    &           &                   & 0.003  & 0.0001   & ...     & 0.0030 & 0.0053 & 0.03507 \\
42  &AK\,IV-287 & 67411989910269696 & 11.553 & 1808     & 0.396   & 0.229  & ...    & 6020.17 & SB1 & 1, 3  \\
    &           &                   & 0.003  & 26       & ...     & 0.064  & ...    & 1648.27 \\
43  & HII\,522  & 65194584193758336 & 11.684 & 23.8375  & 7.10    & 0.109  & ...    & 3.59702 & SB1 & 1, 3  \\
    &           &                   & 0.003  & 0.0005   & 1.05    & 0.023  & ...    & 0.54151 \\
44  & HII\,1348 & 66734720809017856 & 12.229 & 94.805   & 9.874   & 0.5543 & 0.7799 & 32.7101 & SB2 & 1, 3 \\
    &           &                   & 0.003  & 0.012    & 0.118   & 0.0017 & 0.0098 & 0.4432  \\
45  & HCG\,384  & 66937859881182848 & 12.782 & 542.11   & 2.352   & 0.6624 & 0.820  & 1239.42 & SB2 & 1, 3 \\
    &           &                   & 0.003  & 0.27     & ...     & 0.0080 & 0.035  & 68.44  \\
46  & HII\,3104 & 65660158649542784 & 12.847 & 1312.5   & 3.941   & 0.207  & ...    & 419.421 & SB1 & 1, 3  \\
    &           &                   & 0.003  & 4.5      & 0.926   & 0.051  & ...    & 107.608 \\
47  & HII\,1653 & 66816187748666624 & 12.904 & 548.2    & 0.737   & 0      & ...    & 743.826 & SB1 & 1, 3  \\
    &           &                   & 0.003  & 1.1      & 0.227   & ...    & ....   & 229.107 \\
48  & HCG\,76   & 64456262134709376 & 15.732 & 32.747   & 1.979   & 0.1328 & 0.917  & 18.3020 & EB  & 3, 13 \\
    &           &                   & 0.003  & 0.002    & 0.263   & 0.0043 & 0.013  & 2.4349  \\
\enddata
\tablecomments{The second line for each system lists the 1$\sigma$ uncertainties. References in the last column are: (1)--(6) are the same as in Table~\ref{t1}; (7) \cite{Torres2025}; (8) \cite{Torres2020}; (9) \cite{Fritzewski2025}; (10) \cite{Gao2025}; (11) \cite{Torres2020b}; (12) \cite{Gaia2023}; (13) \cite{David2016}.}
\label{t2}
\end{deluxetable*}
}

\software{Astropy \citep{Astropy2013, Astropy2018, Astropy2022}, PETAR \citep{Wang2020, Wang2022}, MCLUSTER \citep{Kupper2011, Wang2018}, YBC \citep{Chen2019}
          }

\bibliography{sample701}{}
\bibliographystyle{aasjournalv7}
\end{CJK*}
\end{document}